%                       TeX File
%----------------------------------------------------------------------
%\magnification=1100
\overfullrule=0pt
\baselineskip=20pt
\parskip=0pt
\def\dag{\dagger}
\def\del{\partial}
\def\ybh{\hat{\bar\y}}
\def\yh{\hat{\y}}

\def\a{\alpha}             
\def\b{\beta}              
     \def\G{\mit\Gamma}   
\def\d{\delta}        
\def\e{\epsilon}

\def\q{\theta}

\def\m{\mu}

\def\p{\pi}        \def\P{\Pi}

\def\y{\psi}            
\def\w{\omega}     \def\W{\mit\Omega}   
\def\br{\langle}
\def\ke{\rangle}
\def\ve{\vert}

\def\px{{\hat \p }^{x}}
\def\py{{\hat \p }^{y}}
\def\ph{\hat \p }
\def\phd{{\hat \p }^{\dag }}
\def\phdd{{\hat \p }^{\dag 2}}
\def\phdp{{\hat \p }^{\dag p}}
\def\hx{\hat X }
\def\hy{\hat Y }
\def\hz{\hat Z }
\def\hzb{\hat {\bar Z }}
\def\hzs{\hat z}
\def\hzbs{\hat {\bar z}}
\def\pz{{\hat \P}^0}
{\settabs 5 \columns
\+&&&&UB-ECM-PF 96/1\cr}
\bigskip
\centerline{\bf MULTIPLE EDGES OF A QUANTUM HALL SYSTEM }
\centerline{\bf IN A STRONG ELECTRIC FIELD }
\bigskip\bigskip
\centerline{Rashmi Ray$^1$ and Joan Soto$^2$}
\bigskip
 
\centerline{ Departament d'Estructura i Constituents de la Mat\`eria}
\centerline{ Facultat de Fisica}
\centerline{Universitat de Barcelona}
\centerline{Diagonal, 647}
\centerline{E-08028-Barcelona-Catalonia-Spain}
\bigskip
\centerline{\bf Abstract}
In this article we show that if the  electrons in a quantum Hall 
sample are subjected to a constant electric field in the plane
of the material, comparable in magnitude to the background 
magnetic field on the system of electrons, a multiplicity of edge 
states localised at different regions of space is produced in the 
sample. The actions governing the dynamics of these edge states are 
obtained starting from the well-known Schr\"odinger field theory for a 
system of non-relativistic electrons, where on top of the
constant background 
electric and magnetic fields, the electrons are further subject to 
slowly varying weak electromagnetic fields. In the regions between 
the edges, dubbed as the \lq \lq bulk", the fermions can be integrated out 
entirely and the dynamics expressed in terms of a local effective action 
involving the slowly varying electromagnetic potentials. It is further shown 
how the \lq \lq bulk" action is gauge non-invariant in a particular way and 
how the edge states conspire to restore the $U(1)$ electromagnetic
gauge invariance of the 
system.

\bigskip 
\vfill 
\noindent{$^1$  rashmi@greta.ecm.ub.es} \ \ 
\noindent{$^2$  soto@ecm.ub.es}

\vfill \eject
{\bf I. Introduction}
\bigskip
Field theories of planar systems have enjoyed a considerable measure of
popularity in recent years, particularly in the sphere of condensed
matter physics.  Specifically, the field theory of the Quantum Hall 
Effect (QHE), in its
various incarnations, is the focus of a great deal of vigorous activity.
As it was noted by several authors [1,2,3,4], a finite Hall sample supports
gapless excitations at the edge (edge states) which have been
experimentally measured. The action governing the edge excitations plays
a crucial role in maintaining the electromagnetic $U(1)$ gauge invariance
[2,3,4]
and it would be a worthwhile endeavour to extract this action in a clean
fashion from the underlying microscopic action describing the generic
QHE.

  \bigskip

In the sequel, we have devoloped a technique which allows us
to calculate the effective action for strong magnetic and
electric fields in a QH sample. A key ingredient is the proper
splitting of the microscopic action into two pieces, one leading to a
local effective action in
the bulk and the other to the edge states. Perturbation theory must be
organised in a different fashion for the two pieces. The maintenance of
the $U(1)$ gauge invariance emerges in a rather non-trivial fashion.

\bigskip

We begin with a system of planar non-relativistic electrons in a
constant strong magnetic field perpendicular to the plane and a constant
electric field in the plane. The electric field need not be small.
In fact we shall see that if the electric field is comparable to the
magnetic field an infinite number of edges are created even when the
chemical potential is small, as opposed to the case of a perturbatively
weak electric field where only one edge is produced [5]. This is a novel
feature of this work.

\bigskip

Another new feature of this work is that we have developed an operator
technique which permits a gauge invariant treatment of the strong
magnetic and electric fields. The Landau gauge and the symmetric gauge,
for instance, which are among the popular choices for the strong
magnetic field, can be easily seen to lead to the same operator algebra.
This is also true for any interpolating gauge choice between them.
It is important to notice that the splitting of the electromagnetic
field between the constant strong magnetic and electric components and
the slowly varying (weak) background depends on the gauge choice for the
former. Although it is indeed technically possible to work within a
given
gauge with its explicit single-particle wave functions, the gauge
invariance of the slowly varying background will be lost in the
effective action unless a great deal of caution is exercised. The gauge
independent operator algebra obviates this problem. It further avoids
using explicit single-particle electronic wave functions, which
simplifies the calculations immeasurably.

\bigskip

Another interesting discussion included here concerns the role of
the electron mass. Since we are working in a non-relativistic theory for
the electrons it is implicit that $m^2>>B$, and hence any term
suppressed
by $1/m$ should be considered smaller than the same term suppressed by
$1/\sqrt{B}$ . However, it is quite easy to realize the
scenario $B>>m^2$. Although a proper treatment of this case
would require the full relativistic theory, our calculation exhibits the
flexibility required to adapt to this new relationship between the external
parameters. Namely, finite mass effects can be incorporated
systematically.

\bigskip

Last but not least, we have provided what we believe to be a very clear
derivation of the edge actions. We also discuss in detail the
ambiguities inherent to anomalous gauge theories and how they can be
accomodated to obtain a perfectly gauge invariant effective action once
the contributions from the bulk have been taken into account.

\bigskip

Having established the salient features of this work, we go on to its
organizational details. In the first section, we establish the notation
and discuss the operator algebra for the unperturbed system together
with the classification of the perturbations in powers of the large
background magnetic field. We further discuss the role of the chemical
potential (which specifies the ground state of the system) in defining
the edge of the system in the presence of a constant electric field. We
note that near the edges, the derivative expansion that we have
envisioned is bound to fail. We indicate how to excise small
neighbourhoods of the edges and reserve them for a separate treatment,
thereby establishing a clear domain of validity for the derivative
expansion. In the second section, we provide a detailed example of the
derivative expansion we have performed and indicate how interesting
terms emerge with a minimum effort. The third section contains the final
results of the derivative expansion in the bulk, where we establish how
potentially damaging terms cancel out in a clean fashion. The net effect
of this expansion is to provide a local electromagnetic effective action
governing the bulk of the system. This effective action is not gauge
invariant in itself but all the gauge non-invariance is pushed to the
edges. In the fourth section we show how to deal with the neighbourhoods
of the edges in a direct fashion. This results in the emergence of 1+1
dimensional anomalous chiral gauge theories at the edges. We show that
local counterterms can be added to the basic anomaly to fully cancel
the gauge dependence  of the bulk action obtained in the previous
section. The final section is reserved for indicating further avenues of
work along these lines. Non-trivial manipulations required in the
calculation are displayed in the appendix.

%\bigskip
\vfill\eject
\centerline{\bf II.  Notation and Formulation.}
\bigskip
The system we wish to discuss is described by the Lagrangian density
$${\cal L} = \hat {\bar {\y }} (\vec x, t)\bigl[iD_0 +{1\over{2m}}{\vec D}^2 
\bigr]\hat \y (\vec x,t) \eqno (2.1) .$$
Here $\yh (\vec x, t)$ is the second quantized fermion field operator 
satisfying the standard anticommutation relation
$\{ \yh (\vec x,t), \ybh (\vec y, t) \} = {\d }^{(3)} (\vec x - \vec y )$.
Here 
$$ iD_0 \equiv i\del _{t} -a^0 (\vec x,t)-
A^0(\vec x, t)
\eqno (2.2)$$
and
$$ -i\vec D \equiv -i\vec \nabla -\vec A(\vec x,t) -\vec a(\vec x,t) 
\eqno(2.3) .$$
The uniform background fields $E$ and $B$ arise from  $A_0$ and $\vec A$. 
$E$ is
in the plane and taken to be along the x-direction and the 
magnetic field
$B$ is perpendicular to the plane. They are taken to be of comparable 
magnitude.
The action corresponding to (2.1) has the obvious U(1) gauge invariance
$$\yh (\vec x,t) \rightarrow \exp {i\q (\vec x,t)}\yh (\vec x,t)$$
$$ \d [a^{\m }(\vec x,t)+A^{\m }(\vec x,t)] = \partial ^{\m } \q (\vec x,t).$$
In the sequel, we shall not display the dependence of the functions on 
$\vec x$ and $t$ explicitly.
This action may be written in a more transparent form
$$S = \int d \vec x dt \ybh (\vec x,t) \bigl [\pz -h \bigr]\yh (\vec 
x,t) \eqno (2.4),$$
where $h$ is the corresponding single-particle hamiltonian,
$$ h = {1\over{2m}}{(-i\vec \nabla -\vec A -\vec a)}^2+ Ex
+a^0 
\eqno (2.5) $$
and
$${\ph }^0 = i\del_{t}-A^0+Ex . $$
Integrating the fermions out one obtains an effective action in terms of the
$a^{\m }$.
$$S_{\rm eff}= -i {\rm Tr } \log \bigl [ {\ph }^0 -h \bigr ] \eqno (2.6). $$
To evaluate $S_{\rm eff}$ exactly, we need to know the exact spectrum of
$h$, which is not possible as the $a^{\m }$ are arbitrary. However,
$$h_0 \equiv {1\over{2m}}{(-i\vec \nabla -\vec A)}^2 +
Ex 
\eqno(2.7) $$
has a well-known spectrum. We thus obtain the effective action 
by perturbing around the eigen-basis of $h_0$, with ${1\over{B}}$ as the 
small parameter. It should be remarked that the actual parameter is the
ratio of any gauge invariant local configuration of the dimensions of
$[mass]^2$ constructed out of $a^{\m }$, with $B$. In the following we
have this in mind when we mention the order in $1/B$ of any given term.

We set up a gauge-independent algebraic procedure for extracting the 
spectrum of $h_0$. 
We take 
$$A^0 \equiv E{\hat x} \a $$
$$A^{x} \equiv (1-\a )E
\hat t + \b B {\hat y}$$
$$A^{y} \equiv -(1-\b )B {\hat x} $$
where it is clear that these potentials lead to the correct background 
fields irrespective of the arbitrary parameters $\a $ and $\b $.
We further define the single-particle operators
$${\hat \P }^0 \equiv i\del_{t}-A^{0}+E {\hat x} \eqno (2.8)$$
$${\hat \p }^{x} \equiv
-i\partial_{x} -A^{x} \eqno (2.9)$$
$${\hat \p }^{y} \equiv 
-i\partial_{y} -A^{y}+m{{E}\over{B}} \eqno (2.10).$$
The associated commutator
$$[\px ,\py ] = -iB \eqno (2.11) $$
is seen to be independent of the arbitrary parameters $\a $ and $\b $.
We further define the single-particle operators
$$\hat {X^{0}} \equiv \hat {t} \eqno (2.12) $$
$$\hat X \equiv \hat x -{1\over{B}}\hat {\p^{y}} \eqno (2.13) $$
$$\hat Y \equiv \hat y +{1\over{B}}\hat {\p^{x}} \eqno (2.14) $$
with the associated commutators
$$[\hat X,\hat Y] ={{i}\over{B}} \eqno (2.15) $$
$$[\hat {\P^{0}}, \hat {X^{0}}] = i \eqno (2.16) .$$
These commutators are also seen to be independent of $\a $ and $\b $.
$\hat X $ and $\hat Y $ are referred to in the literature as guiding
center coordinates.
We can also see quite easily that 
$$[\px , \hx ] = [\py , \hx ] = [\px , \hy ] = [\py , \hy ] = 0 
\eqno(2.17) $$ 
Further,
$$[\pz , \px ]=[\pz , \py ]=[\pz , \hat X ]=[\pz , \hat Y ]=0 $$
and
$$[{\hx }^0,\px ]=[{\hx }^0,\py ]=[{\hx }^0,\hx ]=[{\hx }^0,\hy ]=0. $$
These relations are also insensitive to the choice of $\a $
and $\b $.  They further indicate that the spectrum of the $\hat {\P^{i}}$
(the cyclotron motion) is independent of that of $\hat X$ and $\hat Y$ 
(the guiding center motion). 

We can now construct holomorphic and anti-holomorphic combinations of 
these operators.
$$\ph \equiv \px - i \py \eqno(2.18) $$
$$\phd \equiv \px + i \py \eqno (2.19) $$
$$\hz \equiv \hx + i \hy \eqno (2.20) $$
$$\hzb \equiv \hx - i \hy \eqno (2.21) $$
and the corresponding commutation relations are
$$[\ph , \phd ] = 2B \eqno (2.22) $$
and
$$[\hz, \hzb ] = {2\over{B}} \eqno (2.23) .$$
With these operators, $h_0$ can be re-expressed as
$$h_0 = {1\over{2m}}{\phd } {\ph }  + E \hat {X} 
+{{\w_c }\over{2}} + {{m}\over{2}}{{\bigl( }{{E}\over{B}}{\bigr) }}^2 \eqno 
(2.24) $$ $\w_{c} \equiv {{B}\over{m}}$ is the cyclotron frequency.
It is easily seen that $h_0$ is diagonalised by the basis
$\{ \vert n,X \ke \} $, where
$$\ph  \ve n,X \ke = \sqrt{2Bn}\ve n-1,X \ke \eqno (2.25) $$
$${\phd }{\ph }\ve n,X \ke = 2Bn \ve n,X \ke \eqno 
(2.26) $$
$$\hx  \ve n,X \ke = X \ve n,X \ke \eqno (2.27) .$$
$$h_0 \ve n,X \ke = E_{\rm n}(X) \ve n,X \ke \eqno(2.28) ,$$
with $$E_{\rm n}(X) \equiv n\w_{c} +EX 
+{{\w_{c}}\over{2}} + {{m}\over{2}}{({{E}\over{B}})}^2
\eqno (2.29). $$
The index $n$ labels the Landau levels. In the event that $E = 0$, $X$
measures the degeneracy of each Landau level (L.L.). The electric field 
lifts this degeneracy. It is important to note that the unperturbed
single-particle spectrum is unbounded below.
In the second-quantised many-body problem, however, the ground state will 
be specified as the one where all the single-particle 
states are occupied up to a given 
energy. This is done by an appropriate choice of the 
chemical potential.
For definiteness, we shall take the chemical potential such that all the
energy levels up to $E_0(0)$ are filled up. This is equivalent to redefining
the origin of energies by dropping the
constant term
${{\w_{c}}\over{2}} + {{m}\over{2}}{({{E}\over{B}})}^2 $ in (2.29), which
we shall do in the rest of the paper. Hence
$$E_{n}(X)\longrightarrow E_{n}(X)=nw_{c} +EX $$
For a given $n$, the shifted energy becomes negative when 
$$X \leq -X_{n} \equiv 
-{{\w_{c}}\over{E}}n \eqno(2.29a). $$
Furthermore, for $\ve \vec E \ve \ll B $, only $X_0 = 0$ remains finite.
All other $X_{n}$ for $n\neq 0$ go to minus infinity.
Thus, in this case the filling of all negative energy single-particle 
states is tantamount to filling only the negative energy states 
corresponding to $n=0$. Thus the system exists only upto $X=0$, which for 
$n=0$ is the same as $x=0$. Hence for $n=0$, the system exhibits an edge 
in real space at $x=0$[5].
For $\ve \vec E \ve \simeq B$, however, all L.L. contribute to the ground 
state. In this case, an edge develops for each $X_{n}$.

An interesting observation that has to be made is that away from $X=X_{n}$,
the size of               
 $<n,X\vert h \vert n,X>$ 
 is roughly 
$E_{\rm n}(X)$,
which is $O(B)$. However, in the neighbourhood of $X \sim X_{n}$,
since
$E_{\rm n}(X)\sim 0$ the size of
 $<n,X\vert h \vert n,X>$ 
depends on the slowly varying electromagnetic potentials and turns out 
to be $O(1)$.
 Thus near the edges given by $X_{n}$, the perturbation 
theory about $h_0$ fails. To ensure its validity
, we excise $\e $ neighbourhoods of the $X_{n}$ out from the 
integral over $X$, which arises when we calculate the trace (2.6) in the 
$\vert n, X>$ basis,
 and treat 
the fermion modes in these neighbourhoods separately. For the remaining 
modes, the perturbation is obviously valid. These fermion modes that are 
treated separately, lead to the edge states of the system.

Having discussed these finer points, let us now go on to organise the 
perturbative expansion.

The perturbations on $h_0$ are given by
$$V \equiv h-h_0 =a^0  + {1\over{2m}}\bigl [{\vec a}^2 -\vec a 
\cdot (-i\vec \nabla - \vec A) - (-i\vec \nabla - \vec A ) \cdot \vec a \bigr ] 
\eqno(2.30). $$
We define the holomorphic and the anti-holomorphic combinations of the
components of the vector potential.
Namely,
$$A \equiv a^{x} + i a^{y}$$
and 
$$\bar A \equiv a^{x} - i a^{y}. $$
Further,
$$ z\equiv x+iy $$
and
$$ \bar z \equiv x-iy .$$
We note that 
$$ [\ph , A(\hat z, \bar {\hat z})] = -2i\del_{z}A \eqno(2.31) $$
and
$$ [\bar A(\hat z, \bar {\hat z}), \phd ] = 2i\del_{\bar z}{\bar A} 
\eqno (2.32). $$
Here, we have used the relations 
$$\hat z = \hat Z - {{i}\over{B}}\phd $$
$$\hzbs = \hzb + {{i}\over{B}}\ph \eqno(2.33) $$
that are obtained from (2.13), (2.14), (2.18)-(2.21).  
In terms of these quantities, the perturbation (2.30), is written as
$$V=a^0+{{E}\over{B}}a^{y}-{1\over{2m}}b+{1\over{2m}}{\vec a}^2-
{1\over{2m}}(A\ph +\phd \bar A ).$$
Here $b=\del_x a‡^{y}- \del_y a^{x}$. 
Furthermore, any analytic function 
$$f(\hat z, \bar {\hat z})= \sum_{p,q} {1\over{p!\ 
q!}}{({{-i}\over{B}})}^{p}{({{i}\over{B}})}^{q} {\phdp }{\ph }^{q} 
\sharp {\del_{Z}}^{p} {\del_{\bar Z}}^{q} f(\hz , 
\hzb ) \sharp . $$
Here the symbol $\sharp \ \ \sharp $ symply indicates indicates the 
anti-normal
ordering forced onto $\hz , \hzb $ due to the normal ordering 
of $\ph , \phd $. 
Using these, (2.30) may be rewritten and expanded in powers of $1/B$ as
$$V = V^{(1/2)} + V^{(0)} + V^{(-1/2)} + V^{(-1)} + \cdots \eqno(2.34) $$
where
$$ V^{(1/2)} = -{1\over{2m}}(A \ph + \phd \bar A) 
\eqno(2.35) $$
$$V^{(0)} = {\W } - {{i}\over{2mB}}[{\ph }^2 \del_{\bar z}{A} - ({\phd })^2
\del_{z}{\bar A} - {\phd }{\ph }(\del_{z}{A} - \del_{\bar z}{\bar A} )]
\eqno(2.36)$$
$$\eqalignno { V^{(-1/2)} & = {{i}\over{B}}(\ph {\del_{\bar z}} {\W } - 
{\phd } {\del_{z}} {\W } )\cr
& -{1\over{4m{B}^2}}{\phd }{\ph }^2 (2\del_z \del_{\bar 
z}A-{\del_{\bar z}}^2 \bar A)\cr
& -{1\over{4m{B}^2}}{\phdd } {\ph }(2\del_{z} \del_{\bar z 
 }\bar A - {\del_{z}}^2 A )& (2.37) \cr }$$ 
$$V^{(-1)} = {1\over{{B}^2}} 
{\phd }{\ph }\del_{z}{\del_{\bar z}} {\W } +{{i}\over{4mB^3}}{\del_z}
{\del_{\bar z}} (\del_z A - \del_{\bar z} {\bar A})\phdd {\hat \p }^2 
\eqno(2.38) .$$
Here $$\W \equiv a^0 + {{E}\over{B}}a^{y} - {1\over{2m}}b + 
{1\over{2m}}{\vec a }^2 \eqno(2.39). $$
Here all the functions are 
anti-normal ordered with respect to $\hz , \hzb $ even though the anti-
normalisation symbol has not been displayed explicitly. In (2.37), (2.38),
we have omitted off-diagonal terms that do not contribute to the
order we are working at.
Re-inserting all this back into the expression for the effective action given
in (2.6), expanding the logarithm and retaining contributions upto $O(1/B^2)$
we obtain
$$\eqalignno{
S_{eff}&= 
i {\rm Tr } \bigl[  - \log D +\sum_{i=1}^{6} L_{i}\bigl]\cr
&\cr
L_1&=  {1\over{D}}(V^{(0)}+V^{(-1)})\cr 
L_2&=  {1\over{2}}{1\over{D}}V^{(1/2)}{1\over{D}}V^{(1/2)}\cr 
L_3&=  {1\over{D}}V^{(1/2)}{1\over{D}}V^{(-1/2)}\cr 
L_4&=  {1\over{2}}{1\over{D}}V^{(0)}{1\over{D}}V^{(0)}\cr 
L_5&=  {1\over{D}}V^{(1/2)}{1\over{D}}V^{(1/2)}{1\over{D}}V^{(0)}\cr 
L_6&=  {1\over{D}}V^{(1/2)}{1\over{D}}V^{(1/2)}{1\over{D}}
    V^{(1/2)}{1\over{D}}V^{(1/2)}  & (2.40)\cr }$$
where $D \equiv {\hat\P}_{0}-h_{0}$. The first term in the trace in 
(2.40), $-\log D$, does not involve 
the perturbative electromagnetic potentials. It is just the vacuum energy 
and we are not interested in it in the present article.
The trace is realised in terms of the tensor product of
the basis of $h_0$ given earlier with the basis of eigenfunctions of
 $\Pi^0$. Indeed, since $[{\hat \P}^0 ,h_0]=0$ the eigenvectors of $D$ read
$$D \ve n,X,w \ke = \G_{n}(X,w) \ve n,X,w \ke \eqno (2.41) .$$
$$\G_{n}(X,w) = \w -
E_{n}(X) $$
such that $\ve n,X,w\ke=\ve n,X\ke\ve w\ke$ where
$${\hat \P}^0\ve w\ke=w\ve w\ke \eqno (2.42)$$
 
The specification of the ground state of the system, namely that all the 
negative-energy single-particle states are filled means that the integral 
over the frequency $w$ must be defined such that
$$\int {{d\w }\over{2\p }}{1\over{\G_{n}(X,w)}} = i \q (-E_{n}(X)) \eqno 
(2.43). $$
Furthermore, as we have mentioned earlier, we have to excise $\e $ 
neighbourhoods of $X = -X_{n}$ to ensure the validity of the perturbation 
theory. This has the following direct consequence. Consider the 
derivative of (2.43) with respect to $E_n(X)$
$$\int {{d\w }\over{2\p }} {1\over{{\G_{n}}^2(X,w)}} = -i\d (E_{n}(X)) 
\eqno (2.44). $$
 Since, for a given n, $E_{n}(X)=0$ only when $X=-X_{n}$, the 
delta function never contributes as $X = -X_{n}$ is not permitted 
according to the
%the regularisation of the $X$ integral
criteria adopted above to 
define a valid perturbation theory. Hence, coincident denominators 
can be dropped owing to the exclusion of the $\e $ neighbourhoods.
We shall do so consistently in the sequel.
\bigskip
%\vfill\eject
\centerline{\bf III. A Sample Calculation.}
\bigskip
In this section, we provide a non-trivial instance of the perturbative 
calculation outlined in the preceding section actually being performed.
We choose to work out the term $L_3$ in equation (2.40) in 
detail.
Realising the trace operation as mentioned earlier, this term is
$$L_3 = \sum_{n=0}^{\infty }\int dX \int {{d\w }\over{2\p }} 
\br n,X,\w \ve 
{1\over{D}} V^{(1/2)}{1\over{D}}V^{(-1/2)} \ve n,X,\w \ke \eqno (3.1).$$
We now use the action of $\ph $ and $\phd $ on the basis states from
(2.25) and its complex conjugate and that of $D$ from (2.41) to obtain

$$ \eqalignno {L_3 & = \cr
& {{1}\over{2m}}\sum_{n=0}^{\infty }\int dX \int 
{{d\w }\over{2\p }} \Bigl(
{1\over{{\G_{n}}{\G_{n+1}}}}
\br X,n,\w \ve 
(2i(n+1))
\sharp A\sharp \sharp
\del_{z}{\W }\sharp
 + ({{2n(n+1)}\over{m}})\sharp A \sharp\sharp ({\del_{z}}{\del_{\bar z 
}}{\bar A} - {1\over 2}{\del_{z}}^2 A )\sharp \ve X,n,\w \ke \cr 
& + {1\over{\G_{n}}{\G_{n-1}}}\br X,n,\w \ve (-2in)
\sharp \bar {A} \sharp\sharp\del_{\bar z}{\W }\sharp 
+({{2n(n-1)}\over{m}}){\sharp \bar {A} \sharp
} \sharp ({\del_{z}}{\del_{\bar z}}A-{1\over 
2}{\del_{\bar z}}^2 {\bar A}) \sharp \ve X,n,\w \ke
\Bigr) &(3.2) \cr }$$
We have written $\G_{n}$ as a shorthand for $\G_{n}(X,w)$.
We have also used that the commutator of $1/D$ with any function $f(\hat Z,
\hat {\bar Z}, t)$ is suppressed by $1/B$ (see (A.9)) and hence it does not
contribute to $L_3$ at the order we are considering. However, there are 
contributions to $L_2$ emanating from such a commutator.
We can now use the properties (2.25) and (2.26) to bring the Landau level
index $n$ in the states to zero with no other alteration in (3.2).
We next introduce the identity $\int dt \vert t \rangle \langle t \vert$
in the $\vert w \rangle$ subspace. Since $\langle t\vert w\rangle = e^{-i
wt}$ all the $w$-dependence in the states disappears and it only remains in 
$\G_{n}$. We have
$$ \eqalignno {L_3 & = \cr 
& {{1}\over{2m}}\sum_{n=0}^{\infty }\int dX \int 
{{d\w }\over{2\p }}\int dt  \Bigl( {1\over{{\G_{n}}{\G_{n+1}}}}
\br X,0\ve 
(2i(n+1))
\sharp A\sharp \sharp
\del_{z}{\W }\sharp
 + ({{2n(n+1)}\over{m}})\sharp A \sharp\sharp ({\del_{z}}{\del_{\bar z 
}}{\bar A} - {1\over 2}{\del_{z}}^2 A )\sharp \ve X,0 \ke \cr 
& + {1\over{\G_{n}}{\G_{n-1}}}\br X,0 \ve (-2in)
\sharp \bar {A} \sharp\sharp\del_{\bar z}{\W }\sharp 
+({{2n(n-1)}\over{m}}){\sharp \bar {A} \sharp
} \sharp ({\del_{z}}{\del_{\bar z}}A-{1\over 
2}{\del_{\bar z}}^2 {\bar A}) \sharp \ve X,0 \ke 
\Bigr) &(3.3) \cr }$$
We use $\ve X\ke$ for $\ve X,0\ke$ in the following.

At this point we recall that all the functions of $\hz $ and $\hzb $ above
are separately anti-normal ordered.
However, as displayed in the appendix in (A.6), the product of two 
individually anti-normal ordered quantities can itself be re-expressed as
an infinite series of anti-normal ordered quantities, more and more
subleading in (1/B). At the order we are working only the first term in
the series contributes to $L_3$. However the next to leading term in the 
series contributes to $L_2$. Hence we can just substitute the separate
anti-normal orderings in (3.3) by a global anti-normal ordering which we 
will not explicitely write down (e.g. for the first term in (3.3) we have
$\sharp  A\sharp\sharp\partial_{z} \Omega\sharp \sim
\sharp  A\partial_{z} \Omega\sharp \longrightarrow
 A\partial_{z} \Omega $).
We then use equations (A.10), (A.11) from the appendix to disentangle the 
multiple
denominators in (3.2).
%Namely,
%$${1\over{{\G_{n}}{\G_{n+1}}}} = {1\over{\w_{c}}}({1\over{\G_{n+1}}}-
%{1\over{\G_{n}}})$$
%and
%$${1\over{{\G_{n}}{\G_{n-1}}}} = {1\over{\w_{c}}}({1\over{\G_{n}}}-
%{1\over{\G_{n-1}}})$$
This yields
$$ \eqalignno {L_3 = {{1}\over{2m}}\sum_{n=0}^{\infty }\int dX \int dt
\int {{d\w }\over{2\p }}\bigl[ & {{2im}\over{B}}(n+1)({1\over{\G_{n+1}}}
- {1\over{\G_{n}}}) \br X \ve A \del_{z}{\W }\ve X \ke \cr
& + {2\over{B}}n(n+1)({1\over{\G_{n+1}}} - {1\over{\G_{n}}})
\br X \ve A (\del_{z} \del_{\bar z}{\bar A }- {1\over2}{\del_{z}}^2 A )\ve X 
\ke \cr 
& + {{(-2im)}\over{B}}n({1\over{\G_{n}}}- {1\over{\G_{n-1}}})
\br X \ve {\bar A}{\del_{\bar z}}{\W }\ve X \ke \cr 
& + {2\over{B}}n(n-1)({1\over{\G_{n}}}-{1\over{\G_{n-1}}})
\br X \ve {\bar A}({\del_{z}}{\del_{\bar z}}A-{1\over2}{\del_{\bar z}}^2
{\bar A})\ve X \ke \bigr] & (3.4) \cr }$$
which can be further simplified using (A.12)-(A.15). 
Thus we reduce all the denominators to the same form $\G_{n}$. This enables
us to use (2.43) to integrate over $\w $ and obtain the $\q $ function
displayed therein. This means that from (3.4) we obtain
$$\eqalignno {L_3 = {{1}\over{2m}}\sum_{n=0}^{\infty }\int dX \ \int dt
\ & \bigl[ {{2m}\over{B}}\q (-E_{n}(X)) \br X \ve (A \del_{z}{\W }
-{\bar A}\del_{\bar z}{\W }\ve X \ke \cr
& - {{4in}\over{B}}\q (-E_{n}(X))\br X \ve A(\del_{z} \del_{\bar z}{\bar A}
-{1\over2}{\del_{z}}^2 A )+ {\bar A}(\del_{z} \del_{\bar z}A-{1\over2}
{\del_{\bar z}}^2 {\bar A})\ve X \ke \bigr ] & (3.5) \cr }$$
In its turn, the $\q $ function is used to yield an $n$-dependent upper
bound on the $X$ integral. Namely, the $X$ integral goes from $-\infty $ to
$-X_{n}$.
 Furthermore, as shown in (A.5), the integral over $X$ 
of a function
that is anti-normal 
ordered and is evaluated in the $X$ basis can be easily re-expressed
as the integral over real space of a c-number valued function that can 
be readily obtained from the original anti-normal ordered q-number valued 
function.
In this integral over real space, the $x$ integral ranges from $-\infty $ to
$-X_{n}$, a consequence of the restricted range of integration of $X$. The
$y$ integral is on the other hand over the entire $y$-axis. 
Thus, going through
 this last step
 we have for $L_3$,
$$ \eqalignno {L_3=-{{i}\over{2\p }}\sum_{n=0}^{\infty }\int_{-\infty 
}^{-X_{n}}dx \int dy \int dt \bigl [ &  
b(a^0 + 
{{E}\over{B}}a^{y} - {1\over{2m}}b + {1\over{2m}}{\vec 
a}^2)\cr 
& + {{2n}\over{m}}(A\del_{z}\del_{\bar z}\bar A + \bar {A} 
\del_{z}\del_{\bar z}A -{1\over 2}A {\del_{z}}^2 A-{1\over 2}{\bar 
A}{\del_{\bar z}}^2{\bar A}) \bigr ] \cr 
& +{{i}\over{2 \p }}\sum_{n=0}^{\infty }\int dy \int dt [a^0 a^{y} - 
{1\over{2m}}a^{y}(b-{\vec a}^2)+{{E}\over{B}}({a^{y}})^2]\ve_{x=-X_{n}} 
&(3.6) \cr } $$
We note here that apart from terms that obviously live in the \lq \lq bulk" 
there are terms that live at the edges of the bulk. These terms have
arisen as a consequence of re-expressing some terms in the bulk
in a convenient form
through judicious integrations by parts.
Notice also that we have taken the $\e$-neighbourhoods to zero at the end
of the calculation.

% and are the resulting surface terms 
%in the x-direction.
%These are local terms in the effective action, expressible in terms of 
%perturbative 
%gauge potentials which act only at the boundaries of the sample. Thus we
%have re-expressed $L_3$ in a very convenient form. We perform very similar
%manipulations 
%for the other terms in (2.40).
% In some of the terms, an additionalfeature
% has to be taken into account to obtain all possible terms to the same
%order in 1/B. For instance, while simplifying $L_2$, a very important
%contribution arises from the fact that since $D$ contains $\hat X$, it does
%not commute with functions of $\hz $ and $\hzb $. As shown in the appendix
%in (A.14), the terms arising due to this are typically proportional to
%$E/B$. 

\bigskip
%\vfill\eject
{\bf IV. Results for the bulk}
\bigskip

Now that the perturbation theory in the bulk has been scrupulously
defined and a significant example of a calculation has been shown in
section 3, we proceed to display the results for the various
contributions in (2.40).

\bigskip

Notice that the first term in (2.40) is just an irrelevant vacuum 
contribution 
that we drop. Let us organise the remaining contributions as follows
$$ S_{eff}^{bulk}=i
\sum_{n=0}^{\infty}\int dx\int dy \int dt \int {dw\over
2\pi}\int
%{X_{n}+\epsilon}
%^{X_{n+1}}
%-\epsilon}
%\
 dX
\vert \langle X,0,w\vert x,y,t\rangle \vert^2
{1\over \Gamma_{n}(X,w) }
\sum_{i=1}^{6} L_{i}
% \Bigl(
%L_{1}+L({1\over 2},{1\over 2})+L({1\over 2},-{1\over 2})
%+ L(0,0)+L({1\over 2},{1\over 2},0)
%+L({1\over 2},{1\over 2},{1\over 2},{1\over 2})
% \bigr)
 \eqno (4.1)  $$

We obtain:

$$ \eqalignno{
 L_{1}
 =&
%-{b\over 2m} +
{\W } - {{n}\over{m}}b
%-{n^2\over mB}
%\partial_{\bar z}\partial_{z}
%b
+ {2n\over B}
\partial_{\bar z}\partial_{z}
{\W }
-{{n(n-1)}\over{mB}}\del_z \del_{\bar z}b 
   &(4.2) \cr
L_{2}
 =&
-{1\over 2m} A\bar A
 +{1\over 2mB}(
\partial_{\bar z} A\partial_{z}\bar A
+\partial_{z} A\partial_{\bar z}\bar A
)
& \cr &
-{1\over 4B}\Bigl[ -i\bar A\partial_{t} A +iA\partial_{t}\bar A
-{E\over B}\bigl(
\bar A ( \partial_{z} A -\partial_{\bar z} A)
-A (\partial_{z} \bar A -\partial_{\bar z} \bar A )  \bigr)\Bigr]
&(4.3) \cr
L_{3}
 =&{i\over B}\bigl[
%-i\bar A\partial_{\bar z}b+2mi
\bar A\partial_{\bar
z}
{\W } 
%i A\partial_{z}b-2mi
- A\partial_{
z}
{\W }
 \bigr]
& \cr &
-{2n\over mB}\bigl[
 \bar A (
-{1\over 2}
\partial_{\bar z} \partial_{\bar z}\bar A
+\partial_{\bar z} \partial_{z} A )
+ A (
-{1\over 2}
\partial_{ z} \partial_{z}\bar A
+\partial_{\bar z} \partial_{z} \bar A )  \bigr]
   &(4.4) \cr
 L_{4}
 =& -{2n+1\over mB}
\partial_{z} \bar A\partial_{\bar z} A
   &(4.5) \cr
L_{5}
 &=
{i\over 2mB}(
-\bar A \bar A\partial_{\bar z} A
+ A A\partial_{z} \bar A
+\bar A  A\partial_{z} A
-\bar A A\partial_{\bar z} \bar A  )
   &(4.6) \cr
L_{6}
 &=0
   &(4.7) \cr
 }         $$
%where
%$$
%\quad b:=i(
%\partial_{\bar z}\bar A- \partial_{z} A )
%\eqno (4.8)
%$$

A few comments are in order. The second and third terms in the first
line of (4.3) arise from the re-normal ordering of the first term
(see (A.6)). The second line in (4.3) arises from the fact that $D$
and $V^{(1/2)}$ do not commute (see (A.9)).

 A few cancellations
can be readily seen from (4.2)-(4.7). The term ${1\over 2m} A\bar A $
 from (4.3) cancels against the same term in
% $ \tilde a_0  $
$\W$
 from (4.2). (4.6) cancels against the terms with
${1\over 2m} A\bar A $ in
% $ \tilde a_0  $
$\W$
 from (4.4).
As we have indicated in (2.43), the integral over $\w $ gives rise to a
step function. This step function, in turn, provides a finite upper limit
to the $X$ integral. We have shown in (A.1)-(A.5) how this translates into a 
finite upper bound for the $x$ integral. 
Therefore partial
integration with respect to $x$ gives rise to boundary terms which must
be kept.
Let us divide the final result into bulk and boundary terms. In each
contribution there are both mass independent and mass dependent terms.
We also rewrite $A$ and $\bar A $ in terms of $a^{x}$ and $a^{y}$ and
$\del_{z}$ and $\del_{\bar z}$ in terms of $\del_{x}$ and $\del_{y}$.
Consider first the bulk effective action. 
We get 
$$
 S^{bulk}={1\over {2\p }}
\sum_{n=0}^{\infty}\int_{-\infty}^{-X_{n}}dx\int dy \int dt
(L^{top}+L^{m}) \eqno (4.9) $$
$$\eqalignno{ L^{top}=&-{1\over 2} \epsilon^{\mu \nu
\rho}a_{\mu}\partial_{\nu}a_{\rho}- 
B(a_0+{E\over B}a_{y})
&(4.10)\cr
L^{m}=& -{2n+1\over 2m}b^2 + {{2n+1}\over{2m}}bB &(4.11) }$$
%where
%$x_{n}=X_{c}-{E\over mw^2}+{nw\over E}$ and the limit
%$\epsilon\rightarrow 0$ has been taken.
 The boundary effective action
reads $$
 S^{bound}_{n}= {1\over{2\p }}
%\sum_{n=0}^{\infty}
\int dy \int dt
(L^{top}+L^{m}) \eqno (4.12) $$
$$\eqalignno{ L^{top}=& -{n\over 2}\partial_{x}(
a^0+{E\over B}a^{y}
)-{1\over 2}(a^{y} a^0 +{E\over B}(a^{y})^2)
&(4.13)\cr
L^{m}=&-{{n^2}\over {4mB}}\partial_{x}b+{2n+1\over 4m}\partial_{x}(a^{y})^2
 &(4.14) }$$
Notice that the bulk action is gauge invariant but the boundary action
is not. This is so because on the one hand there are explicit gauge
non-invariant terms in (4.13)-(4.14) and on the other hand the gauge
variation of (4.9) gives rise to boundary contributions due to the 
Chern-Simons term in (4.10). We shall see in
the next section how the edge states fix up the lack of gauge invariance at
the boundary.

\bigskip
{\bf V. Edge States}
\bigskip
In section 2 we excluded from the calculation small neighbourhoods
around $X_{n}$, which correspond to zero modes of the hamiltonian $h_0$.
In these neighbourhoods $h_0$ is $O(B^0)$ instead of $O(B^1)$, and hence
the perturbation theory must be reorganised.
Notice first that
$V^{-{1\over 2}}$ amd $V^{-1}$
are still
perturbations. We shall ignore them for the moment and see later on that
they actually give rise to higher order terms. Let us then split $D$
into its diagonal and off-diagonal parts.

$$\eqalign{D=& O_{d} + \Delta O \cr
           O_{d}= & {\hat \Pi}_0-h_0-a^0-{E\over B}a^{y}
-{1\over 2m}A\bar
A+{1\over 2m}({\pi^{\dagger}\pi\over B} +1)b \cr
           \Delta O=& \Delta O^{1\over 2} +\Delta O^{0} \cr
           \Delta O^{1\over 2}=& {1\over 2m}(
A\pi+\pi^{\dagger} \bar A)\cr
           \Delta O^{0} =& {i\over 2mB}\partial_{\bar z} A 
\pi^2-{i\over 2mB}\partial_{z} \bar A{\pi^{\dagger}}^2 }
\eqno (5.1) $$

The effective action for all these neighbourhoods reads
$$ S^{edges} =
-i \sum_{p=0}^{\infty}
\int_{-X_{p}-\epsilon}^{-X_{p}+\epsilon} dX
\sum_{n=0}^{\infty} \int {dw\over 2\pi} \langle n,X,w\vert \log(
O_{d}+\Delta O) \vert n,X,w\rangle
\eqno (5.2) $$
%$$ X_{p}:={1\over E}[-(p+{1\over 2}){B\over m} +{mE^2\over 2B^2}+\mu ]
%$$
where $X_{p}$ is defined in (2.29a).
Let us next focus on a fixed $p$. Notice that when $\epsilon
\rightarrow 0$ only singular terms give a finite contribution to
$S^{edge}$. Those may occur only when $h_0$ is close to zero.
Otherwise, a
local expansion about $h_0$ which gives rise to piecewise
continuous functions as we have seen in the previous section, is legitimate. 
$h_0$ gets close to zero
 only if $n=p,p\pm 1$ at the order we are interested in.
Consider then the formal expansion

$$ S^{edge\; p}=
-i
\int_{-X_{p}-\epsilon}^{-X_{p}+\epsilon} dX
\sum_{n=0}^{\infty} \int {dw\over 2\pi} \langle n,X,w\vert \bigl( \log
O_{d} +\sum_{r=1}^{\infty}{(-1)^{r+1}\over r}({1\over{O_{d}}} \Delta O)^{r}
\bigr)
\vert n,X,w\rangle
\eqno (5.3)
$$

Let us first focus on the first term in the expansion for $n=p$

$$ \eqalign{
S^{edge\; p}_{0}=&
-i
\int_{-X_{p}-\epsilon}^{-X_{p}+\epsilon} dX
\int {dw\over 2\pi}  \langle p,X,w\vert
\log \bigl({\hat \Pi}_0-h_0-a^0-{E\over B}a^{y}
-{1\over 2m}A\bar
A+{1\over 2m}({\pi^{\dagger}\pi\over B} +1)b  \bigr)
\vert p,X,w\rangle  \cr
=&
-i
\int_{-X_{p}-\epsilon}^{-X_{p}+\epsilon} dX
\int {dw\over 2\pi}  \langle 0,X,w\vert
\log
\Gamma_{p}
%(\hat X,\hat Y)
\vert 0,X,w\rangle
\cr & \cr
\G_{p}:=&\Gamma_{p}({\hat \P}_0,\hat X,\hat Y):=
 {\hat \Pi}_0-
{B\over m}p
%+{1\over 2})-{mE^2\over B^2} +\mu
-E\hat X +f_{p}(\hat
X,\hat Y)
%\vert 0,X,w\rangle 
 \cr &\cr
f_{p}(\hat X,\hat Y):=&
 -a^0-{E\over B}a^{y}+ -{1\over 2m}A\bar
A+{1\over 2m}(2p+1)b }
\eqno (5.4) $$
Let us next insert the identity in the subspace $<\hat Y , t>$ and use
the representation
$$ \hat X=
{i\over B}\partial_{Y}
\quad , \quad {\hat\Pi}_0=i\partial_{t}
\eqno (5.5) $$
We have
$$
S^{edge\; p}_{0}=
-i
\int_{-X_{p}-\epsilon}^{-X_{p}+\epsilon} dX
\int {dw\over 2\pi}
 \int dY \int dt
\langle 0,X,w\vert
 0,Y,t\rangle
\log
\Gamma_{p}(i\partial_{t},
{i\over B}\partial_{Y}
, Y)
%\bigl( i\partial_{t}-{B\over m}(p+{1\over 2})+\mu-i
%{E\over B}\partial_{Y}
% +f_{p}(
%{i\over B}\partial_{Y}
%, Y) \bigr)
\langle 0,Y,t
\vert 0,X,w\rangle
\eqno (5.6)$$

Using the explicit expressions for the wave function
$$
\langle 0,Y,t\vert 0,X,w\rangle=
\sqrt{B\over 2\pi}e^{-iwt}e^{iBXY}
       \eqno (5.7)
$$
and making the shift
$$ X \rightarrow - X_{p}+{k\over B}  \eqno (5.8)$$
we have
$$
S^{edge\; p}_{0}=
-i
%{1\over B}
\int_{-\epsilon B}^{\epsilon B} {dk\over 2\pi}
\int {dw\over 2\pi}\int dt \int dY \int dt
\, \log \bigl(w-{E\over B}k+
i\partial_{t}-i{E
\over B}\partial_{Y}
 +f_{p}(X_{p}
, Y) \bigr)  + O(\epsilon )
\eqno (5.9)$$

For $\epsilon B \rightarrow \infty$ the above is nothing but the
effective
action for a 1+1 dimensional chiral fermion, the lagrangian for which reads
$$ L_{p}^{chf}=\int dt \int dY \psi^{\dagger}\bigl(
i\partial_{t}-i{E
\over B}\partial_{Y}
 -a^0-{E\over B}a^{y}
+{1\over 2m}(2p+1)b
-{1\over 2m}A\bar A
\bigr) \psi
\eqno (5.10) $$

Notice that (5.9) is already $O(B^0)$ and hence any contribution
suppressed by $1/B$ should be consistently neglected. In particular,
that includes contributions from
$V^{(-1/2)}$ amd $V^{(-1)}$
which we have already disregarded.
Formula (5.10) is not yet the right lagrangian for the edge states , but
it contains its essential features, which we would like to comment upon.
First of all there is only one term, namely
$-{1\over 2m}A\bar A $ ,
 which is not gauge
invariant. This
term cancels out with other terms which also contribute to the edge
action, as we will see below, so we shall disregard it in the
following discussion. The remaining terms can be split into  mass
independent and mass dependent terms. The mass independent
terms, which are identical for any edge ($p$ independent) and hence
identical to the ones
obtained for the only edge that exist when the electric field is weak
[5], correspond to the usual lagrangian for a 1+1 chiral fermion (chiral
Schwinger model). The coeficient of the mass dependent term depends
on the particular edge ($p$ dependent) and has not been found before.

It is well known that the chiral Schwinger model is anomalous [6,7]. That
means that not all the symmetries of the classical lagrangian can be
implemented at the quantum level. This is due to UV singularities
%arising in relativistic QFTs
 and implies that the quantum theory is
defined up to local counterterms. The local counterterms encode
the inherent ambiguity of an anomalous gauge theory[8]. They can also be
regarded as a reminder of the possible different results one can
obtain at the quantum level by
using different regularizations. In the relativistic chiral Schwinger
model, one usually requires Lorentz symmetry
at the quantum level. This forces gauge invariance to be spoiled and
reduces the number of possible local counterterms
. In our case the physical situation is quite different. In condensed
matter physics Lorentz symmetry is not supposed to play an important
role. Although in the $m\rightarrow \infty$ limit the edge action
enjoys a $1+1$ Lorentz symmetry (taking $c=E/B$), it is not clear that
it must be considered a fundamental symmetry. In fact, the only
fundamental symmetry that we have in our system is the $U(1)$
electromagnetic gauge invariance. Then the most reasonable criterion to
define quantum mechanically the edge effective action is to require
gauge invariance in the whole system. Since the bulk action contains a
Chern-Simons term which gives rise to the gauge anomaly at the boundary,
we must define the edge action in such a way that cancels out this gauge
anomaly, together with the rest of non-gauge invariant local
counterterms. Recall that the cancellation of the gauge anomaly coming
 from the Chern-Simons term in the bulk cannot be carried out by means
of local counterterms at the edge. It actually requires the existence of
extra degrees of freedom at the edge, in this case the chiral fermion.
 Therefore once we have choosen a regularization for the
edge effective action which cancels the gauge anomaly , the non-gauge
invariant local counterterms are automatically fixed. The only freedom
we have are gauge invariant local counterterms (up to total derivatives in
$t$ and $y$)
at the edge. Let us list
them below
$$
\eqalign{d=1: \quad\quad & a^0 \quad , \quad a^{y} \cr
d=2: \quad\quad & \partial_{x}a^0 \quad , \quad
\partial_{x}a^{y} \cr
d=3: \quad\quad & \partial_{x}^2a^0 \quad , \quad
\partial_{x}^2a^{y}
}
\eqno (5.11)
$$
The coeficients of these terms contribute to the observables and must be
fitted experimentally or calculated numerically from the fundamental
theory.

One may wonder at this point as to how it comes about that even though we 
started with a well-defined problem we end up with a few free parameters 
that we cannot calculate. The answer is that for the bulk calculation and for
the edge calculation the assumptions on $\epsilon$, the cut-off which
separates edge and bulk, are different. For the perturbation theory
at the bulk to be correct one needs $E\epsilon >>{\sqrt{B}a^{i}\over
m}
\, ,\,
{a^{i}a^{j} \over m}
\, ,\,
a^0
$
, whereas for the derivation of the edge states $E\epsilon <<1$ is
understood. Therefore there is a region, presumably small, in which none
of the expansions hold. What is remarkable, however, that our ignorance about
this region can be summarized into only a few gauge invariant local
counterterms at the edges.

Let us next analyse the remaining terms in the expansion (5.3) for $n=p$
.
Recall first that
 $1/O_{d}$ is $O(B^0)$ at the edge but $O(B^1)$ elsewhere and that
$\Delta O$ contains a piece $O(B^{1\over 2})$ and a piece $O(B^0)$.
 We notice that only a very particular class of terms gives
contributions of the same order as (5.4).
 Indeed, if
 $1/O_{d}$
is at the
edge,
$\Delta O$
 brings the following
 $1/O_{d}$
out of the edge.The next
$\Delta O$
may either keep the next
 $1/O_{d}$
out of the edge or bring it back to the edge. In the first case we
obtain a term which is suppressed at least by a power of $1/B$ with
respect to (5.4).
In the second case we  obtain a contribution of the same order as (5.4)
 due to
${\Delta O}^{1\over 2}$
 whereas
${\Delta O}^{0}$ gives rise to contributions suppressed by $1/B$
. Therefore
we must add up all the terms corresponding to the second case in which
only
${\Delta O}^{1\over 2}$
is involved
. We have
$$ \eqalign{
S^{edge\; p}_{0^{\prime}}=
&
-i
\int_{-X_{p}-\epsilon}^{-X_{p}+\epsilon} dX
\int {dw\over 2\pi}
\sum_{j=1}^{\infty}-{1\over 2j}
\langle p,X,w\vert
({1\over O_{d}}\Delta O {1\over O_{d}}\Delta O )^{j}
\vert p,X,w\rangle
\cr
=&
-i
\int_{-X_{p}-\epsilon}^{-X_{p}+\epsilon} dX
\int {dw\over 2\pi}
\sum_{j=1}^{\infty}-{1\over 2j}
\langle p,X,w\vert
({1\over \Gamma_{p}} [{
1\over \Gamma_{p-1}}{pB A\bar A \over 2m^2} +
{1\over \Gamma_{p-1}}{(p+1)B A\bar A \over 2m^2}
])^{j}
\vert p,X,w\rangle \cr
\sim &
-i
\int_{-X_{p}-\epsilon}^{-X_{p}+\epsilon} dX
\int {dw\over 2\pi}
\sum_{j=1}^{\infty}-{1\over 2j}
\langle 0,X,w\vert
(-{1\over \Gamma_{p}}
{A\bar A \over 2m}
)^{j}
\vert 0,X,w\rangle }
\eqno (5.12)
$$
The same kind of terms leading to order $B^0$ also appear for $n=p\pm 1$
. We have
$$\eqalign{
S^{edge\; p}_{\pm}=
&
-i
\int_{-X_{p}-\epsilon}^{-X_{p}+\epsilon} dX
\int {dw\over 2\pi}
\sum_{j=0}^{\infty}-{1\over 2(j+1)}
\langle p\pm 1,X,w\vert
{1\over O_{d}}\Delta O
({1\over O_{d}}\Delta O {1\over O_{d}}\Delta O )^{j}
 {1\over O_{d}}\Delta O
\vert p\pm 1,X,w\rangle
\cr
%=&
%\sum_{j=1}^{\infty}-{1\over 2j}
%\langle p,X,w\vert
%({1\over \Gamma_{n}} [{
%1\over \Gamma_{n-1}}{nB A\bar A \over 2m^2} +
%1\over \Gamma_{n-1}}{(n+1)B A\bar A \over 2m^2}
%)^{j}
%\vert p,X,w\rangle \cr
\sim &
-i
\int_{-X_{p}-\epsilon}^{-X_{p}+\epsilon} dX
\int {dw\over 2\pi}
\sum_{j=0}^{\infty}-{1\over 2(j+1)}
\langle 0,X,w\vert
({(-1)^{1\pm 1\over 2}(p+{1\pm 1\over 2})\over \Gamma_{p}}
{A\bar A \over 2m}
)
(-{1\over \Gamma_{p}}
{A\bar A \over 2m}
)^{j}
\vert 0,X,w\rangle }
\eqno (5.13)
$$
A few technical comments are in order. $\Gamma_{p}$ above must be
understood as the
 $(\hat X, \hat Y )$
-valued operator defined in (5.4).
$A$ and $\bar A$
are also functions of
 $(\hat X, \hat Y )$.
Although
 $\hat X$ and $\hat Y $ do not commute, its commutator is
 $O(1/B)$
.
Hence we have commuted
$A$ and $\bar A$
at will.
The commutator of $1/\Gamma_{p\pm 1}$ with
$A$ and $\bar A$  is also
 $O(1/B)$
and has also been neglected. However,
the commutator of $1/\Gamma_{p}$ with
$A$ and $\bar A$  is
 $O(B^0)$
and must be kept.
Instead, the cyclic property of the trace has been used in (5.13) in
order to bring the first
$A$ and $\bar A$
together.

If we add (5.12) and (5.13) to (5.4) we see that the only effect on the
latter is the cancellation of the gauge non invariant term, $-{1\over{2m}}
{\bar A}A $, 
as claimed
above. We finally have
$$ L_{p}^{chf}=\int dt \int dY \psi^{\dagger}\bigl(
i\partial_{t}-i{E
\over B}\partial_{Y}
 -a^0-{E\over B}a^{y}
+{1\over 2m}(2p+1)b
\bigr) \psi
\eqno (5.14) $$

\bigskip
\centerline{\bf VI. Discussion and Conclusions}
\bigskip
In this article, we have focussed on the rather interesting phenomenon 
that in an arbitrarily strong electric field along the plane, a system of 
planar electrons exhibits a multiplicitly of edges, as opposed to the 
case of a weak electric field, where only one edge, due to the lowest 
Landau level, is seen. 

Apart from this generalisation, the calculation here exhibits a number of 
novel features. First, we have shown that we do not have to make a 
specific gauge choice for the background electric and magnetic fields. In 
fact, we have set up a gauge independent algebra for the set of operators 
that diagonalise the unperturbed hamiltonian $h_0$. This is to be 
contrasted with most calculations in the literature where a gauge choice 
(either Landau or symmetric) is made at the onset for the background 
fields. Moreover, at no point in the calculation are we called upon to 
use the explicit wave functions of the higher Landau levels. 
Only at the very end of the calculation are we called upon the explicit
wave functions of the lowest Landau level, in order to make contact with
the real spatial coordinates.
Most of the 
manipulations are at the operator level and integrals have to be 
performed only at the ultimate stages. This alleviates the tedium of a 
derivative expansion considerably. Furthermore, the only expansion has 
been with respect to the large background magnetic field. The electron 
mass has been throughout retained as an arbitrary parameter and hence 
finite mass effects have already been incorporated.
We have further seen that to set up a viable perturbation theory, an $\e $
neighbourhood of the edge due to each Landau level has to be excised and 
the fermion modes in these intervals treated separately. These modes 
cannot be integrated out and reexpressed through local terms involving 
the perturbative gauge potentials. We have dealt with these separately to 
obtain the chiral edge fermionic actions. The remaining fermionic modes 
can be integrated out and reexpressed in terms of a local effective 
action involving the gauge potentials. This is the "bulk" action. Thus a 
clear separation of the bulk and the edge has been effected in this 
calculation. The bulk action is expectedly not gauge invariant by itself. 
The edge fermionic systems also possess the well-known $U(1)$ gauge 
anomaly. The basic anomaly is however seen to be insufficient to 
compensate for the non-invariance of the bulk. We also need to include 
local counter-terms constructed out the operators in the edge fermionic 
action to render the total effective description gauge invariant. While 
this procedure is quite familiar to aficionados of the theory of 
anomalies, it is instructive to encounter it in a more everyday condensed 
matter context.

Turning from enumerating the virtues of our calculation, we outline 
future as well as ongoing projects in the general direction.
We are currently involved in generalising to the case of relativistic 
fermions, in which case the electronic spin is introduced in a
natural manner. One of the objectives is to see whether the spin could
become an important degree of freedom even in a strong magnetic field
[9,10].
Another issue that will bear closer scrutiny is the effect of the Coulomb
interaction between the electrons and the effect of it on the edge states
of the system[11,12,13]. 
The effect of a strong electric field in the case of the fractional Hall 
effect can also be discussed by looking at the integral Hall effect for 
Jain's \lq \lq super\rq \rq fermions[14]. Work in this direction is under 
way.
   
\bigskip
\bigskip
\centerline{\bf Acknowledgements}
\bigskip
R.R. wishes to thank Prof. B. Sakita for discussions that led to the inception
of this project and Dr. M. Lavelle for interesting comments on gauge 
invariance and its manifestations. He
% also
% thanks the
%
% what about the following?
acknowledges financial support from a
 \lq \lq beca postdoctoral para extrangeros
\rq \rq (modalidad B) of the Spanish MEC. J.S. thanks Prof. G.
Veneziano and the CERN theory group for their hospitality while this
paper was written up. This work is partially supported by 
 the grants 
CICYT AEN95-0590 and
GRQ93-1047.
% for financial support.
\bigskip
\centerline{\bf Appendix }
\bigskip
In this appendix, we shall provide some of the conputational details omitted
in the main body of the paper.

As we have noticed in section III, we have to translate results obtained
in the $\{ \ve n, X, \w \ke \}$ basis to the space-time basis. For this,
as we shall see below, we need $\ve \br \vec x \ve 0,X \ke {\ve }^2 $.
This may be worked out by going to a specific gauge. Here, we choose to
work in the
$$\vec A \equiv (0,-Bx) \eqno(A.1)$$
gauge.
In this gauge, we have
$$\br \vec x \ve 0,X \ke = ({{B}\over{\p }})^{1\over 4}
\sqrt{{{B}\over{2\p }}} e^{-iB(X+mE/B^2)y} e^{-{{B}\over 2}
(x-X{)}^2}\eqno(A.2). $$
In fact, in section III, we are required to evaluate
$\int_{-\infty }^{-X_n}dX\ \ve \br \vec x \ve 0,X \ke {\ve }^2$. With the
wavefunction from (A.2)
we obtain
$$\int_{-\infty }^{-X_n}dX\ \ve \br \vec x \ve 0,X \ke {\ve }^2
={{B}\over{2\p }}Erf(-{\sqrt B}(x+X_n))\eqno(A.3),$$
where
$Erf(x)\equiv {1\over{\sqrt{\p }}}\int_{-\infty }^{x}\ dy e^{-y^2}$.
In the limit of a strong magnetic field,
we get, from the value of the error function as its argument goes to
either plus or minus infinity,
$$\int_{-\infty }^{-X_n}dX\ \ve \br \vec x \ve 0,X \ke {\ve }^2
\longrightarrow {{B}\over{2\p }}\q (-(x+X_n))\eqno(A.4).$$
This means that on performing the $X$ integral, the upper limit on the
integral due to the specification of the ground state, devolves on to the
integral over the spatial coordinate $x$.  
\bigskip
{\bf From the $\ve 0,X, \ke $ basis to space-time functions:}
\bigskip
Next, we indicate the simplication of the following expression which occurs 
frequently in the text:
$$\int_{-\infty }^{- X_n }\ dX\  \br 0,X \ve \sharp f(\hz , \hzb ) \sharp 
\ve 0,X \ke .$$
This can be rewritten as 
$$\int_{-\infty }^{- X_n }\ dX\  \br 0,X \ve f(\hzs , \hzbs )   
 \ve 0,X \ke $$
in view of the fact that normal ordered products of $\ph $ and $\phd $ give
zero matrix elements in the lowest Landau level(recall(2.33)).
Inserting the identity operator in terms of the spatial basis,
namely $\int d\vec x \ve \vec x \ke \br \vec x \ve = {\rm I} $ into the above
expression, we get
$$\int d\vec x f(x,y) \int_{-\infty }^{-X_n}dX\ \ve \br \vec x \ve 
0,X \ke {\ve }^2.$$
Using (A.4), this yields
$$\int_{-\infty }^{-X_n}dX\ \br 0,X \ve \sharp f(\hz , \hzb ) \sharp 
\ve 0,X \ke = {{B}\over{2\p }}\int_{-\infty }^{-X_n}dx\ \int_{-\infty }^{
\infty }dy\ f(x,y)\eqno(A.5).$$
%\bigskip
\vfill\eject
{\bf Re-antinormal ordering:}
\bigskip
Again, in the text, we come across expressions of the form
$\sharp f \sharp \ \sharp g \sharp $, where each of the functions
$f$ and $g$ are separately anti-normal ordered in $\hz $ and $\hzb $.
However, this can be written again in a suitably anti-normal ordered form
by commuting $\hz $ across $\hzb $ wherever they do not occur in 
anti-normal
order in the product $\sharp f \sharp \sharp g \sharp $. As $\hz $ and $\hzb $
are canonically conjugate (upto factors), this leads to an infinite series
of anti-normal functions of increasingly higher order in 1/B.
Namely,
$$\sharp f \sharp \ \sharp g \sharp = \sharp fg \sharp -{1\over B}
\sharp \del_{\bar z}f \del_{z}g \sharp +\cdots \eqno(A.6), $$
where the $\cdots $ indicate terms of higher order in $(1/B)$ that 
have been dropped.
Hence, using (A.5) and (A.6), we have,
$$\int_{-\infty }^{-X_n}\ dX \br X \ve \sharp f \sharp \ \sharp g \sharp 
\ve X \ke = {{B}\over{2\p }}\int_{-\infty }^{-X_n}\ dx 
\int_{-\infty }^{\infty }\ dy [fg - {1\over B} 
\del_{\bar z}f \del_{z}g +\cdots ]
\eqno(A.7). $$
\bigskip
{\bf Commuting $1/D$ with $f(\hat X,\hat Y, \hat t)$ :}
%{\bf $[D,f(\hat X , \hat Y,\hat t )]$:}
\bigskip
Furthermore, we also encounter expressions like
$$\br n,X,w \ve \sharp f \sharp {1\over{D}} \sharp g \sharp 
\ve n,X,w \ke . $$
$D$ does not commute with $f$ or $g$ as it contains $\hat X $ and 
$\hat \P_0$. 
However, using
$$[ D , f ] = -i {{E}\over{B}} \del_{y} f 
+[{\hat \P}_0 ,f]
\eqno(A.8), $$
where 
$[{\hat \P}_0 ,f]=i\partial_{t}f$ in the space-time basis,
we have
$$\br n,X,w \ve \sharp f \sharp {1\over{D}} \sharp g \sharp 
\ve n,X,w \ke
% \simeq
= \br n,X,w \ve {1\over{D}}\sharp f \sharp 
\sharp g \sharp \ve n,X,w \ke 
%+ {{iE}\over{2B}} 
%\br n,X,w \ve {1\over{{D}^2}} (\sharp f \sharp \sharp \del_{y}g \sharp -
%\sharp \del_{y}f \sharp \sharp g \sharp  ) \ve n,X \ke
+ \br n,X,w \ve {1\over{D}}[D,\sharp f \sharp ]{1\over D}
\sharp g \sharp \ve n,X,w \ke 
 \eqno(A.9). $$
The second term in the rhs is $O(1/B)$ suppressed with respect to the
first term. The formula above can be further iterated until one obtains 
all $1/D$s acting on the states plus higher order terms.
\bigskip
{\bf Multiple Denominators:}
\bigskip
We also have to perform a bit of algebra with the $\G_{n}(X)$ defined
in equation (2.42). The principal results that we note here are
$${1\over{\G_{n} \G_{n+1}}} = {1\over{\w_{c}}}({1\over{\G_{n+1}}}
- {1\over{\G_{n}}}) \eqno(A.10)$$
$${1\over{\G_{n} \G_{n-1}}} = {1\over{\w_{c}}}({1\over{\G_{n}}}
- {1\over{\G_{n-1}}}) \eqno(A.11)$$
Furthermore, we have
$$\sum_{n=0}^{\infty }{{(n+1)}\over{\G_{n+1}}} = \sum_{n=0}^{\infty }
{{n}\over{\G_{n}}} \eqno(A.12) $$
$$\sum_{n=0}^{\infty }{{n}\over{\G_{n-1}}} = \sum_{n=0}^{\infty }
{{n+1}\over{\G_{n}}} \eqno(A.13) $$
$$\sum_{n=0}^{\infty }{{n(n+1)}\over{\G_{n+1}}} = \sum_{n=0}^{\infty }
{{n(n-1)}\over{\G_{n}}} \eqno(A.14) $$
and
$$\sum_{n=0}^{\infty }{{n(n-1)}\over{\G_{n-1}}} = \sum_{n=0}^{\infty }
{{n(n+1)}\over{\G_{n}}} \eqno(A.15). $$

Other results may be obtained as required by repeated applications of (A.10)
and (A.11).
%\vfill\eject
\bigskip
\bigskip
\centerline{\bf References}
\bigskip
\bigskip
\item{[1]}  B.I. Halperin, Phys. Rev. {\bf B 25}, 2185, (1982).
\item{[2]}  X.G. Wen, Phys. Rev. {\bf B 43}, 11025, (1990) and references 
therein.
\item{[3]}  M. Stone, Ann. Phys. (N.Y.), {\bf 207}, 38, (1991); 
Int. J. Mod. Phys. {\bf B5}, 509, (1991).
\item{[4]}  J. Fr\"ohlich \& T. Kerler, Nucl. Phys. {\bf B354}, 365, (1991).
\item{[5]}  R. Ray \& B. Sakita, Ann. Phys. (N.Y.), {\bf 230}, 131, (1993).
\item{[6]}  R. Jackiw \& R. Rajaraman, Phys. Rev. Lett. {\bf 54}, 1219, (1985)
\item{[7]}  L. Faddeev \& S. Shatashvili, Phys. Lett. {\bf 167 B}, 225, (1986)
\item{[8]}  H. Leutwyler, Phys. Lett. {\bf 152 B}, 78, (1985) 
\item{[9]}  M. Stone, \lq The Magnus Force on Skyrmions in Ferromagnets and
Quantum Hall Systems\rq , cond-mat/9512010 and references therein.
\item{[10]}  S.L. Sondhi, A. Karlhede, S.A. Kivelson, E.H. Rezayi, Phys. Rev.
{\bf B 47}, 16419, (1993).
\item{[11]} R. Ray \& G. Gat, Mod. Phys. Lett. {\bf B 8}, 687, (1994) and
referencs therein.
\item{[12]} B. Sakita, \lq\lq electromagnetic interactions of electrons in
the lowest Landau level: an application of $W_{\infty }$ transformations \rq
\rq , (An article dedicated to the memory of Professor R. E. Marshak).
\item{[13]} A. Lopez \& E. Fradkin, Phys. Rev. {\bf B 44}, 5246, (1991).
\item{[14]} J. Jain, Phys. Rev. Lett. {\bf 63}, 199, (1989).  
\end